\documentclass[12pt,preprint]{aastex}
\def\gtorder{\mathrel{\raise.3ex\hbox{$>$}\mkern-14mu
             \lower0.6ex\hbox{$\sim$}}}
\def\ltorder{\mathrel{\raise.3ex\hbox{$<$}\mkern-14mu
             \lower0.6ex\hbox{$\sim$}}}

\def\Msun{\>{\rm M_{\odot}}}

\shorttitle{$^6$Li in Li-Poor Stars with Planets}
\shortauthors{Mandell, Ge, \& Murray}

\begin{document}
\title{A Search for $^6$Li in Lithium-Poor Stars with Planets}
\author{Avi M. Mandell \& Jian Ge}
\affil{Pennsylvania State University}
\affil{Department of Astronomy \& Astrophysics, University Park, PA 16803, USA}
\author{Norm Murray}
\affil{Canadian Institute for Theoretical Astrophysics}
\affil{University of Toronto, Toronto, Ontario M5S 3H8, Canada}

\begin{abstract}
Using high-resolution, high quality spectra we investigate the
presence of $^6$Li in two lithium-poor stars that host extrasolar
planetary systems.  We present improved atomic and molecular line
lists for the region in the vicinity of the lithium line at 6707.8
\AA, and we produce an excellent fit to the solar spectrum.  From line
profile fitting, we find results consistent with no $^6$Li in either
of the lithium-poor planet-bearing stars or in three comparison stars
with and without planets, and 1-$\sigma$ upper limits of 0.04 for the
isotopic ratios of the two lithium-poor stars give an upper limit of
0.3 Jupiter masses of material with primordial abundances that could
have been recently deposited in their outer layers.  Our results
suggest that post-main sequence accretion of planets or planetary
material that is undepleted in lithium is uncommon.
\end{abstract}

\keywords{planetary systems: formation and evolution - stars: abundances
- stars: lithium - line: identification}

\section{Introduction} 
\label{intro_sec}

To explain the presence of massive extrasolar planets very close to
their parent stars, current formation models often incorporate orbital
migration \citep{lin96,mur98} and/or planetary scattering
\citep{wei96,ras96}.  A possible consequence of these processes would
be the ingestion of planets or protoplanetary material into the
atmosphere of the star if migration is not halted.  It is theorized
that this deposition of planetary material could be detected as
excesses of rare elements, most notably the $^6$Li
isotope. Theoretical calculations \citep{for94,pro89} predict that
$^6$Li is completely destroyed in the pre-main sequence phase of
stellar evolution for solar-mass stars, and anomalously high ratios of
$^6$Li to $^7$Li in normal main- sequence stars could be an important
diagnostic of planet formation processes.

However, if stellar pollution does occur in some fraction of planetary
systems, there are additional systematic characteristics which could
influence the measured ratio of $^6$Li to $^7$Li in a star.  According
to stellar theory, the elimination of $^6$Li during formation of a
star is due to exposure to high temperatures at depth from convective
circulation.  Once the convective zone retreats, any remaining lithium
will be slowly depleted over the lifetime of the star.  Therefore the
current abundance ratio depends on the stellar age and the spectral
type (which determines the size of the convection zone).  Accordingly,
the detection probability would increase for younger, more massive
stars and decrease for older, less massive ones.  Also, the
preservation of polluting material from a planet relies on the
accretion of the planet occurring after the convection zone has
retreated from the core, which in turn depends on the planet formation
timescale, the efficiency of migration mechanisms, and the size and
density of the planet that is accreted.  Therefore, in order to
constrain models of migration and the efficiency of absorption of
planetary material into the outer atmosphere of a star, it is
important to analyze a large sample of both lithium-rich and lithium-
poor planet-bearing stars with a range of different ages and
atmospheric characteristics.

\citet{isr01} was the first group to report measurements of the
lithium isotopic abundance ratio in a planet-bearing star.  They
observed HD 82943, a star with several close-in giant planets, as well
as the star HD 91889 for comparison.  Analysis of the 6707 \AA\ lithium
resonance line yielded the predicted result for HD 91889
($^6$Li/$^7$Li$=-0.002\pm0.006$), but for HD 82943 the ratio was
determined to be $^6$Li/$^7$Li$=0.126\pm0.014$.  This interesting
result was challenged by Reddy {\sl et al.} (2002, hereafter R02), who
observed HD 82943 as well as seven other stars hosting extra-solar
planets and six comparison stars with similar atmospheric
parameters. After applying their analysis to the Sun as well as
another planet-bearing star with very low lithium abundance (16 Cyg
B), they decided that additional unidentified metal lines in the
lithium region were necessary for accurate modeling, and subsequently
concluded that no $^6$Li was necessary to accurately model their
observations.  They attributed the discrepancy with \citet{isr01} to
the additional metal lines incorporated in their line analysis.

Israelian {\sl et al.} (2003, hereafter I03) performed a reanalysis of
HD 82943, investigating the role of the unidentified lines by measuring
several stars of different effective temperature to monitor the
strength of the line.  They concluded that the Ti lines used by R02
are only adequate at higher effective temperatures, and decided that
the Si identification by \citet{mul75} was a better fit.  Using Si
instead of Ti, I03 derived a new isotope ratio of
$^6$Li/$^7$Li$=0.05\pm0.02$.

The extensive yet inconclusive line analysis of recent work suggests
that the detection of $^6$Li abundance enhancements in planet-bearing
stars is fraught with obstacles.  Analysis of the 6707 \AA\ lithium
region is difficult to begin with, due to a plethora of contributing
blends from both metal lines and molecular CN.  There are over 40 weak
atomic lines identified within 2 \AA\ around the lithium line, and
there is evidence of many unknown sources of opacity.  The effect of
these lines is minimal when simply measuring the overall lithium
abundance in lithium-rich stars.  However, accurate line
identifications become critical when attempting to detect the minute
contributions from the weak $^6$Li isotope.

Observing stars with moderate to high lithium abundances mitigates
some of these factors: the overall lithium feature and therefore the
$^6$Li asymmetry are stronger, and contributions from weak blending
lines become less important.  However, since the measurement of
anomalies in $^6$Li can only be judged through a measurement of
anomalies in the ratio of $^6$Li to $^7$Li, a larger abundance of
$^7$Li in the star means that the effect of adding $^6$Li through
planet engulfment is reduced.  Therefore an overabundance of $^6$Li
that would change the overall ratio significantly in lithium-poor
stars may not appear in lithium-rich stars.  Conversely, if the
deposition of planetary material into the outer layers of a star is
significant during planetary migration, there may be correlations with
overall lithium abundance and the presence of close-in giant planets
\citep{san02}.  In addition, analysis of weak stellar lines with good
quality data helps to illuminate underlying blending features and
decreases the ambiguity between a valid detection of $^6$Li and an
inaccurate line list.  This point is especially salient since there
are several different line lists postulated for the weak blending
lines around the lithium line.

To this end, we have decided to concentrate on observing several
bright planet- bearing stars with lower lithium abundances at very
high S/N, and we have analyzed them using the various different viable
line lists in order to discern the resulting differences in the
quality of the synthetic profile fit and the final limits on the
isotopic ratios.  We have observed 47 Ursa Majoris and Upsilon
Andromedae, two very bright lithium-poor stars with established
detections of close-in extrasolar planets.  To check our analysis
techniques we have also analyzed HD 209458, a lithium-rich
planet-bearing star that has been observed by R02, as well as several
stars without known planets. We have achieved S/N per pixel of
$\sim1000$ for 47 UMa and $\upsilon$ And and lower S/N for HD 209458 and
the two comparison stars.  We have also created a more detailed line
list than was previously used, especially with regard to the
underlying CN band.  By analyzing the solar spectrum and a carbon arc
spectrum and applying accurate theoretical wavelengths, we have
enhanced the CN line list dramatically, and we have added additional
atomic lines as well.

A summary of observations and a description of the reduction procedure
is given in \S \ref{obs_sec}.  In \S \ref{param_sec} we give a summary
of the sources for the stellar atmospheric parameters used for the
construction of synthetic spectra, and the procedure used to find the
broadening parameters for each of the stars.  In \S \ref{conv_sec} we
discuss the adjustments made to correct for convective line shifts,
and in \S \ref{line_sec} we describe the sources for the atomic and
molecular line lists.  The uncertainties underlying the critical
blending lines in the lithium region are also discussed.  In \S
\ref{meas_sec} and \S \ref{result_sec} we describe the profile-fitting
procedure for the lithium line and the $\chi^2$ analysis used to find
the best-fit lithium isotopic ratio for each star.  Effects from
uncertainties in the convective wavelength shift and abundances of
blending elements are discussed, and upper limits for the isotopic
ratio for each star are calculated.  Implications for the possibility
of pollution of the stars due to planetary migration are discussed in
\S \ref{discuss_sec}.  We conclude with a summary of the observations
and results in \S \ref{concl_sec}.

\section{Observations}
\label{obs_sec}

Observations were taken with the High Resolution Spectrograph at the
Hobby-Eberly Telescope \citep{tul98} over a period of two years. The
HRS is a fiber-coupled echelle spectrograph, using an R-4 echelle
mosaic with cross-dispersing gratings.  The camera images onto a
mosaic of two 2K x 4K CCDs, with a gap between them that spans
$\sim72$ pixels.  The spectrograph was used in high-dispersion mode,
with an average resolving power of $R = \lambda/d\lambda = 115,000$
measured from thorium-argon lines.  In the high-dispersion mode of the
HRS each resolution element is adequately sampled with 2 pixels per
resolution element.  The spectral coverage is between 4076 \AA\ and
7838 \AA.

Images typically had an integration time of 5 to 15 minutes depending
on brightness, and a total integration time between 1 and 2 hours.  A
ThAr lamp was observed before and after each sequence of
exposures. Due to lithium emission lines in the flatfield illumination
lamp, it was necessary to use extremely high quality observations of
rapidly rotating B stars as flat fields, with signal to noise at least
twice that of the target star.  These calibration stars were observed
as near in time to the program stars as possible.  Atmospheric
absorption lines did not pose a problem in the lithium spectral
region.  The reduction process was repeated with the original flat
fields and nearby continuum regions were used to check the quality of
the final spectrum and to make sure no artifacts were introduced.

Reduction was performed using standard IRAF subroutines for
one-dimensional spectra.  Radial velocity variations due to changes in
heliocentric position and the radial component of the Earth's rotation
between observations were corrected within IRAF, and the individual
images were then corrected for additional minute shifts using the
strong nearby Fe lines.  Continuum normalization was performed by
selecting continuum regions in each image without obvious absorption
lines or bad pixels and fitting a high order polynomial to these
sections.  Images were then combined with appropriate rejection
algorithms.  The estimated S/N per pixel for each star is listed in
Table \ref{params}.

\section{Analysis}
\label{anal_sec}

\subsection{Stellar Parameters} 
\label{param_sec}

\begin{figure}
\plotone{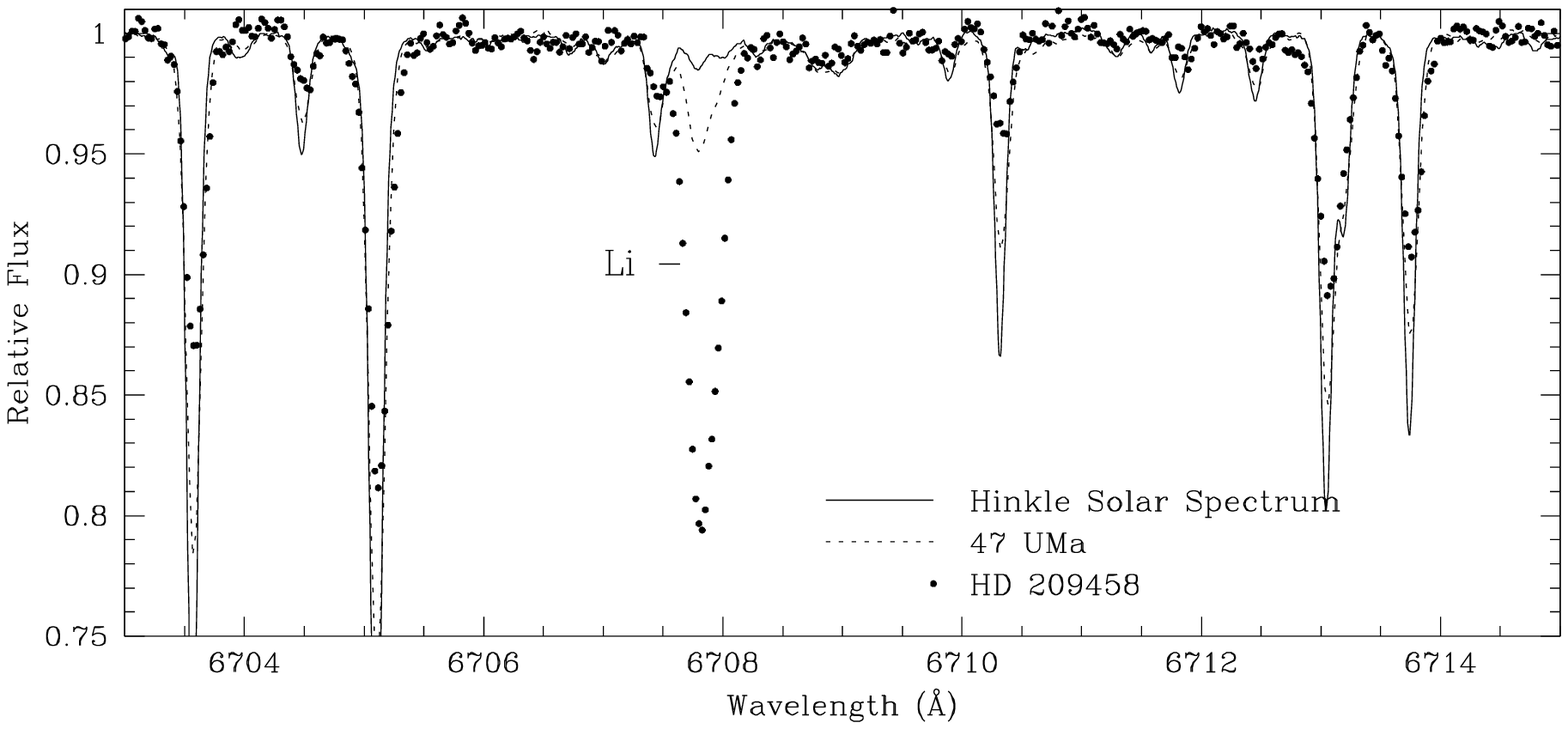}
\caption{Spectra for 47 UMa and HD 209458
overlayed on the spectrum of the Sun from \citet{hin00}.  The spectra
were adjusted to fit the solar spectrum in regions that appeared to be
as close to the continuum as possible. \label{combspec}}
\end{figure}

Atmospheric parameters ($T_{eff}$,log g, and the microturbulence
$\xi_t$) for both the stars with planets and the comparison stars were
taken from the literature (see Table \ref{params} for references).
All of the stars with planets have been well studied, and the
comparison stars are bright nearby stars that also have recent
accurate measurements; uncertainties are on the order of 100K for
$T_{eff}$, 0.1 for log g, and 0.1 for $\xi_t$.  Due to the detailed
nature of earlier analyses of these stars, and the fact that lithium
abundance is relatively insensitive to these parameters, previous
values were considered to be adequate.  Model atmospheres from
\citet{kur95} were used with an updated version of the stellar
analysis code MOOG \citep{sne73} to produce synthetic spectra for each
of the stars. Values for the standard broadening parameters, stellar
rotation ($v \sin i$) and a radial-tangential macroturbulence
parameter ($V_m$), as well as radial velocity for each of the stars
were derived using $\chi^2$ fits to four nearby Fe I lines at 6703
\AA, 6705 \AA, 6713 \AA, and 6715 \AA.  For each line, nearby atomic
lines were found from the VALD-2 database \citep{kup99} and adjusted
to fit the solar spectrum by \citet{hin00}.  For each star the
continuum was also adjusted to match the continuum for the solar
spectrum.  The spectra were initially adjusted to match the solar
spectrum in $\sim$10 regions that appear to be devoid of lines.  These
adjustments were then checked against the synthesized spectrum on
either side of the lithium region, since the list of weak lines in
this region is relatively complete.  The adjusted spectra for the
program stars are shown in Figure \ref{combspec}.

A rough value for the radial velocity (within 0.03 km $s^{-1}$) was
found using estimated broadening parameters.  A three-dimensional grid
was then created by varying the Fe abundance in intervals of 0.02 dex
and the $v \sin i$ and $V_m$ in intervals of 0.02 km $s^{-1}$, for
which the minimum $\chi^2$ value was found.  Best-fit values for the
broadening parameters and the radial velocity were found within 0.01
km $s^{-1}$ for each line.  A mean value for each parameter was then
computed using the three best lines for each star, with an average
uncertainty of 0.2 km $s^{-1}$.  Best fits to the individual Fe lines
varied by as much as 0.5 km $s^{-1}$, most likely due to a combination
of unknown line blends and differences in line shape due to
atmospheric effects (i.e. granulation, convective motion, etc.; see
\citet{asp00} for a detailed discussion).  However, varying the
broadening parameters showed that the computed synthetic spectra are
insensitive to variations in $v \sin i$ and $V_m$ within the $\chi^2$
error limits, as was noted in previous studies.  Final broadening
parameters can be found in Table \ref{results}.

\subsection{Convective Line Shifts}
\label{conv_sec}

\begin{figure}
\plotone{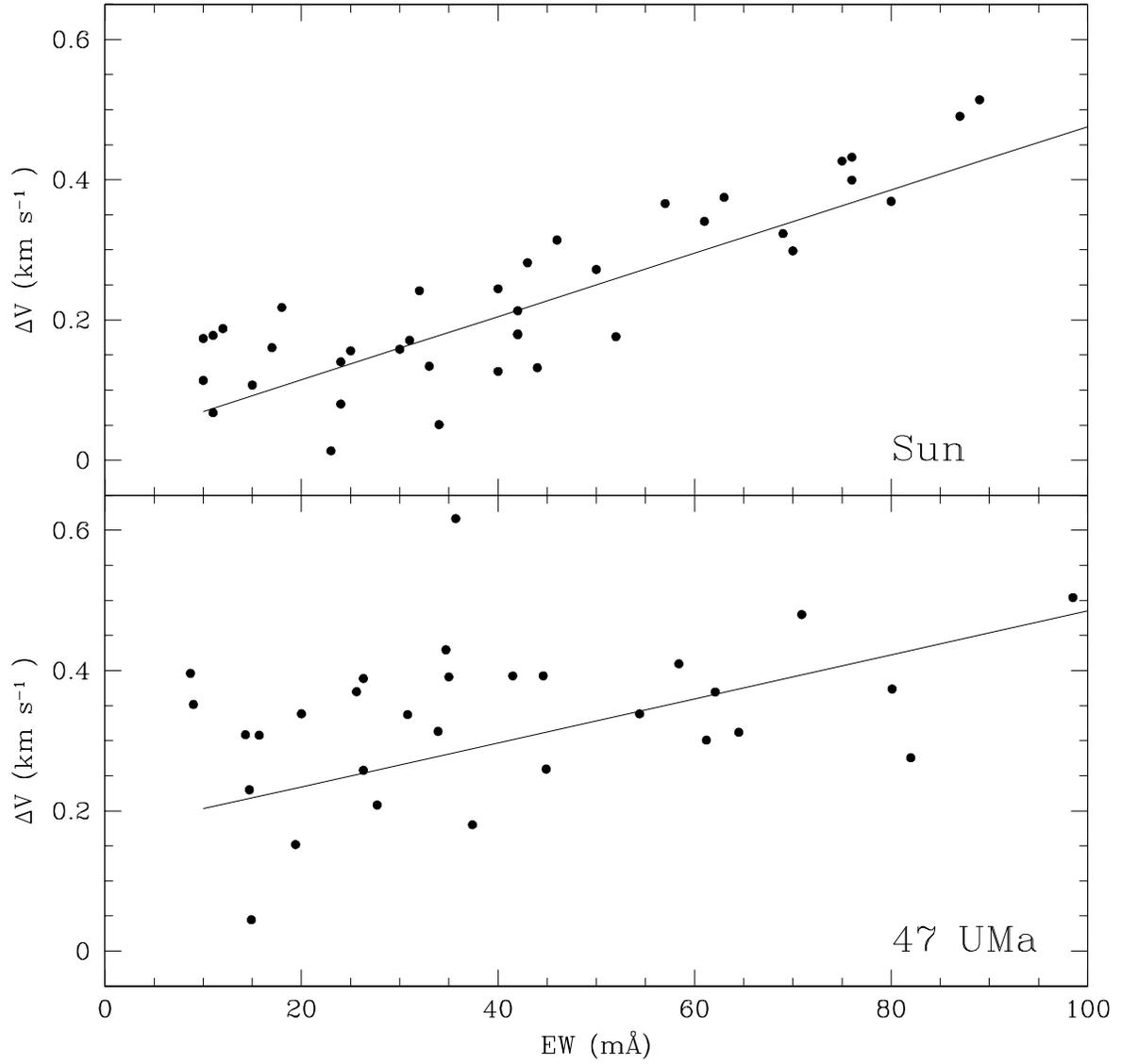}
\caption{Velocity shift of Fe I lines from their
rest wavelengths vs equivalent width for the Sun and 47 UMa due to
differences in convective motion at depth of formation. The measured
relation is in close agreement with \citet{all02} and
\citet{red02}.\label{convsh}}
\end{figure}

In accordance with \citet{all98,all02} and R02, additional radial
velocity shifts due to convective motion were also applied to each
line depending on equivalent width.  Both \citet{all02} and R02
measured comparable correlations between blue shifts in line center
measurements and equivalent width for a large number of Fe lines from
\citet{nav94} for the Sun and other stars of similar spectral type,
noting shifts between -0.2 km $s^{-1}$ and 0.8 km $s^{-1}$.  We have
measured shifts for Fe lines in the solar spectrum of \citet{hin00} as
well as our spectrum of 47 UMa, and we find the same overall
correlations as previous studies (see Figure \ref{convsh}).  However,
the scatter around the mean is quite large, and individual line shifts
often vary by up to 0.5 km $s^{-1}$ from the linear relation.  To
compensate for convective shifts, we have applied an overall
wavelength shift to each line correlated with the equivalent width,
using the relations derived from the solar spectrum by \citet{hin00}.
In addition, wavelength shifts of important lines have been
constrained through fits to the solar spectrum.

\subsection{The Line List}  
\label{line_sec}

The large number of weak blending lines in the 6707 \AA\ region makes
a detailed accurate line list critical to investigations of lithium
abundance.  Accurate laboratory wavelengths and oscillator strengths
have been measured for the $^6$Li and $^7$Li lines themselves
\citep{hob99} but the identifications and properties of many nearby
lines, both atomic and molecular, are quite uncertain.  For this study
we composed an initial line list from the Kurucz atomic database
\citep{kur95} and the Vienna Atomic Line Database (VALD-2,
\citet{kup99}), and additional lines were added from recent studies by
\citet{lam93} and \citet{kin97}.  However, line positions and
oscillator strengths often differ between studies; \citet{kin97}
highlighted several different possibilities for combinations of line
strengths to fit both the solar spectrum and other lithium-poor stars,
as did I03.  An effort was made to include all identified atomic lines
in the lithium region with significant opacity at solar temperatures.
The contribution of molecular CN to the lithium region is also
uncertain.  An early analysis of the CN lines was performed by
\citet{gre68}, who identified eight contributing lines.  Except for
updates in line strengths, this molecular line list has remained
largely unchanged in later studies. Following work by \citet{den91},
we assembled an initial list of lines from work by \citet{jor90}, who
calculated rough line positions and oscillator strengths, with
modifications pointed out by \citet{del98}.  Line positions were
improved when possible using calculations by \citet{kot80} that
incorporate higher order perturbations, and accurate data from
\citet{dav87}.  Following \citet{bra75}, the theoretical data were
then compared with a FTS spectrum of a carbon arc taken from the NSO
digital database \citep{hil03}.  The spectrum was scaled to match the
depth of several unblended CN lines in the solar spectrum at 6702.531,
6703.942, 6704.016, and 6706.733 \AA, and the rotational temperature
estimated from the transition line ratios was found to be similar to
that of the Sun. The line positions and strengths taken from the
literature were then modified to fit the laboratory spectrum; the mean
shift applied was 0.05 \AA, and the mean change in log {\sl gf} was
0.5 dex.  The final CN line list is presented in Table \ref{cn}.  The
line list fits the Li region in the solar spectrum well, recreating
the line asymmetry in the nearby Fe I line and the blend with the
dominant $^7$Li line.

The uncertainties in the immediate region of the Li doublet were
treated using several different line list combinations in order to
exhaust all viable possibilities.  When the preliminary line list was
compiled, it was readily apparent that additional opacity was needed
in the region on the red side of the lithium lines.  The salient
region for lithium-poor stars is the section between 6707.75 and
6708.1 \AA, where the red component of $^7$Li and the stronger blue
$^6$Li component reside.  The solar spectrum by Hinkle shows an
obvious line at 6708.025 \AA, with possible additional opacity sources
at 6707.93, 6707.98, 6708.08 and 6708.1 \AA.  Including the opacity
provided by these blending lines is critical in order to assess the
strength of the $^6$Li lines.  However, there is a notable paucity of
lines listed in the literature for the region beyond 6707.75 \AA.  As
noted by I03, the VALD-2 database contains a weak Ti I line at
6707.964 \AA\ and a V I line at 6708.094 \AA.  There are also many
ionized metal lines in that region, but most of these ionized
transitions have very high excitation values that remove them from
consideration.  Since CN is now removed as a potential source of the
unknown blending lines, we are left to postulate new unknown atomic
lines in order to fit the solar spectrum.

\citet{mul75} first suggested a missing high-excitation Si I line at
6707.025 \AA\ to account for the majority of opacity blueward of
6708.2 \AA, and \citet{kin97} followed this example in their analysis
of various Li-poor solar-type stars.  R02 instead used two Ti I lines
to account for the missing opacity, setting the excitation values at
1.88 eV in order to fit stars ranging in temperature from 5600 K to
6400 K (though the line is overestimated in hot stars).  I03, after
analyzing stars with temperatures from 4500 K to 6200 K, concluded
that Ti I lines greatly overestimate the opacity in cool stars.  They
decided on the 6 eV Si line as the best possibility, but noted that a
low- excitation Ti II line also appears to work well at all
temperatures.  They also noted that the V I lines at 6708.094 and
6708.28 \AA\ used by R02 do not fit stars of non-solar temperature
well.

Since we do not have a large sample of high-resolution spectra at a
variety of temperatures with which to draw our own conclusions, we
must utilize the best options presently available for our stars. After
adjusting the log {\sl gf} values of the basic line list to fit the
solar spectrum, we concluded that additional lines were necessary at
6708.025 and 6708.29 \AA.  For the line at 6708.29 \AA, it appears
that a low-excitation V I line does not fit the width and shape of the
line, and analysis of different stellar temperatures suggests a higher
excitation line would be more appropriate.  For lithium-poor stars the
depth and shape of the line at 6708.29 \AA\ is unaffected by changes
in the lithium line itself, and therefore we can achieve the needed
opacity with any line of the right shape whose abundance we can adjust
with impunity.  We chose to use a Mg I line with an excitation value
similar to nearby neutral Mg lines. After adjusting the log {\sl gf}
to fit the solar spectrum, it appears the shape is adequate to fit all
of our program stars well.

For the line at 6708.025 \AA, a solution is less straight-forward.
From analysis by I03 it appears that both neutral Ti lines and the Si
I line can accurately fit the Sun, but all may overestimate the line
strength in hotter stars and underestimate it in cooler stars (see
Table 2 in I03 for a direct comparison of Si I and Ti I lines).  In
addition, I03 suggests that an ionized Ti line may also be possible;
though they reject it due to the fact the line becomes extremely
weak in hotter stars, the uncertainties in the data make it impossible
to rule out this scenario, and we have decided to include it in our
analyses.  Since the true nature of these blends is still unclear, we
have performed statistical analyses for all three possibilities after
adjusting log {\sl gf} values to fit the solar spectrum.  The final
line list, with the alternative line choices marked, is listed in
Table \ref{atoms}.

\section{$^6$Li/$^7$Li Measurements}   
\label{meas_sec}

The most probable $^6$Li/$^7$Li ratio was determined by performing
$\chi^2$ fits for a variety of $^6$Li/$^7$Li ratios and wavelength
shifts in the lithium lines.  The overall Li abundance was also kept
as a free parameter to allow the line depth to vary.  In addition,
similar fits were performed after varying the abundance for the
element responsible for the unknown blend at 6708.025 \AA, in order to
examine the effects of errors in abundance, excitation energy, or log
{\sl gf} value.

The overall radial velocity was first set using the nearby Fe I
lines. As stated earlier, a linear correction was made to the line
wavelengths in relation to their equivalent widths, but even so
variations between calculated radial velocity values for different
lines ranged between 0.01 km $s^{-1}$ and 0.1 km $s^{-1}$.  This left
a choice to be made between which lines to use for the final value. In
cases where the full line shape of the Fe I line at 6707.45 \AA\ could
be recovered, the line was used to determine the overall radial
velocity, due to the similarities in equivalent width with the target
lines in the lithium region and the fact that this line was used to
calibrate the solar spectrum.  In cases where the Fe line was badly
blended with the Li feature, an average value was calculated from the
four strong nearby Fe I lines.

Abundances for the critical blending lines were taken from the
extensive literature on planet-bearing stars, and then checked using
reference lines near the lithium feature taken from the VALD-2
database (listed in Table \ref{checklines}). In all cases the average
abundance calculated from the reference lines was in good agreement
with the published values.  However, due to inherent line list
uncertainties and the fact that abundances calculated from different
lines vary, separate $\chi^2$ analyses were performed after varying
the abundance of the element used for the 6708.025 \AA\ feature by 0.5
dex.

As noted in previous studies, in all cases it was necessary to shift
the lithium lines towards higher wavelengths in order to fit the blue
side of the $^7$Li line.  The best fit for the Sun was achieved with a
shift of 0.007 \AA, or 0.31 km $s^{-1}$, and remained the same for
each line list variation.  Since the change in wavelength shift with
equivalent width for lithium is unknown, the wavelengths of the
lithium lines were kept as a free parameter in the $\chi^2$ fits.

\section{Results} 
\label{result_sec}

\begin{figure}
\plotone{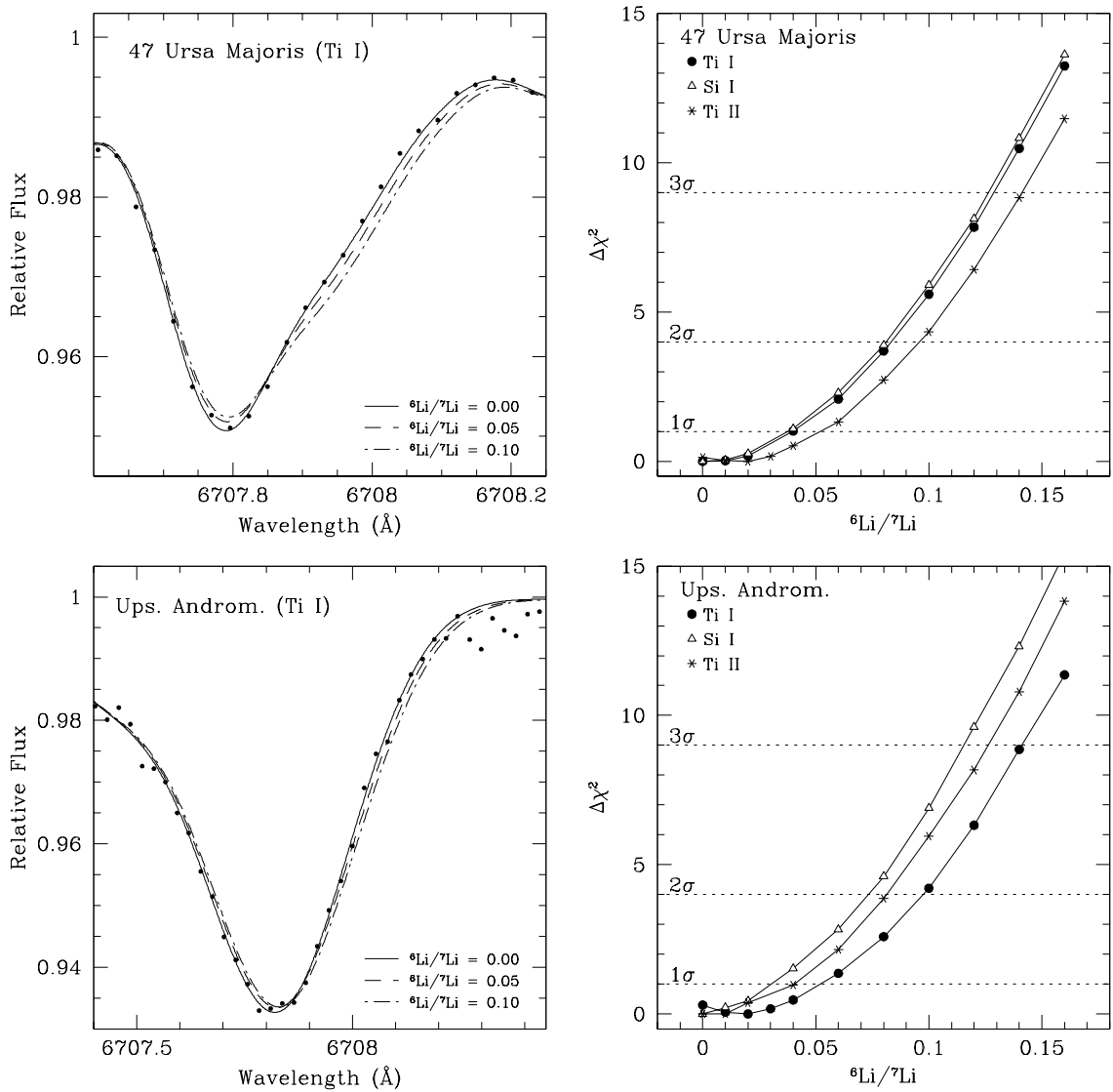}
\caption{Results of $\chi^2$ minimization for 47
UMa and $\upsilon$ And.  The parameter $\Delta\chi^2$ is
calculated by subtracting the best fit from fits where the isotopic
ratio is changed.  The three different analyses are for the three
different choices for the line at 6708.025 \AA.\label{isoplots1}}
\end{figure}

\begin{figure}
\plotone{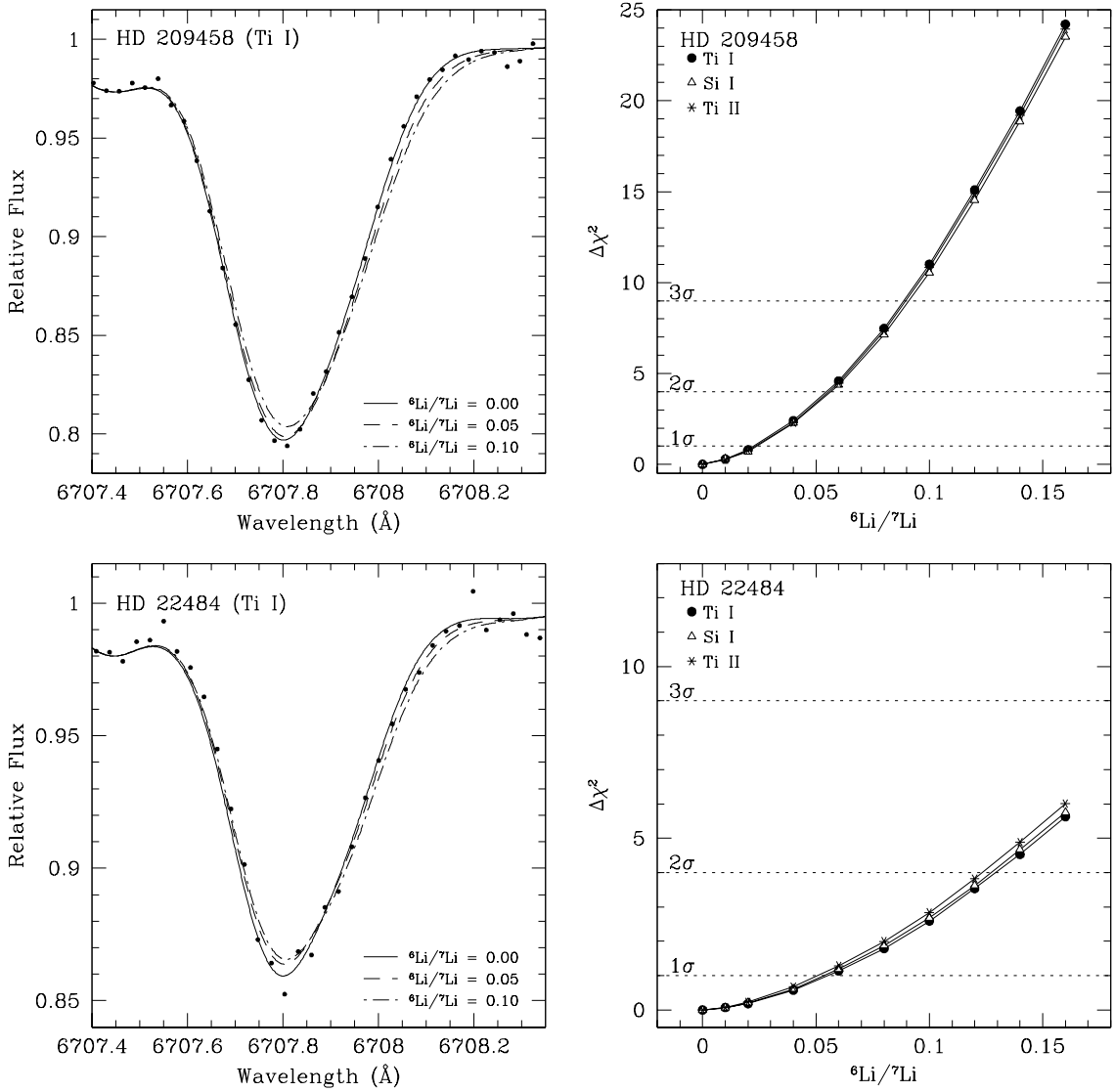}
\caption{Results of $\chi^2$ minimization for HD 209458 and
HD 22484.\label{isoplots2}}
\end{figure}

\begin{figure}
\plotone{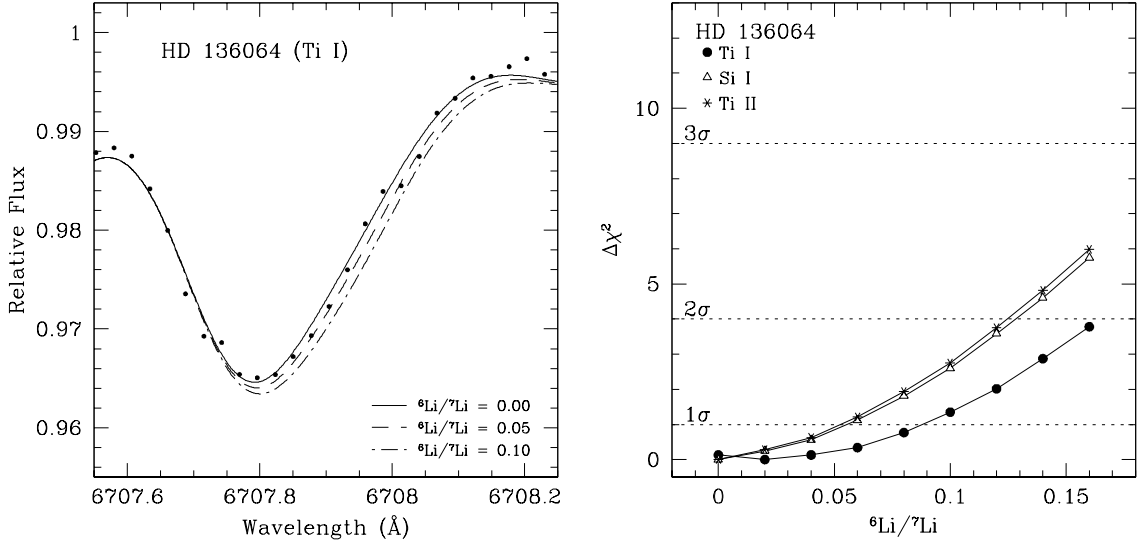}
\caption{Results of $\chi^2$ minimization for
HD 136064.  The S/N for the two references stars HD 22484 and HD 136064
are not high enough to get accurate upper limits, but the fits are
consistent with no $^6$Li.\label{isoplots3}}
\end{figure}

Figures \ref{isoplots1} - \ref{isoplots3} show computed synthetic
profiles for each of the program stars compared with the observed
spectrum, and $\Delta\chi^2$ vs $^6$Li/$^7$Li for the best wavelength
shift and the measured abundance for Ti and Si (where $\Delta\chi^2$
is calculated by varying the lithium isotope ratio and overall
abundance while keeping all other variables constant).  The broadening
parameters, overall lithium abundance, and the best-fit $^6$Li/$^7$Li
ratio for each of the three line list scenarios are listed in Table
\ref{results}.  The errors in both log $\epsilon$(Li) and
$^6$Li/$^7$Li are 1-$\sigma$ errors obtained from the Li profile
fitting, and do not include errors in reference abundances or the
broadening parameters.  The values of $\chi^{2}_{min}$ for the
analyses were between 0.9 and 1.3 for all the program stars.

\begin{figure}
\plotone{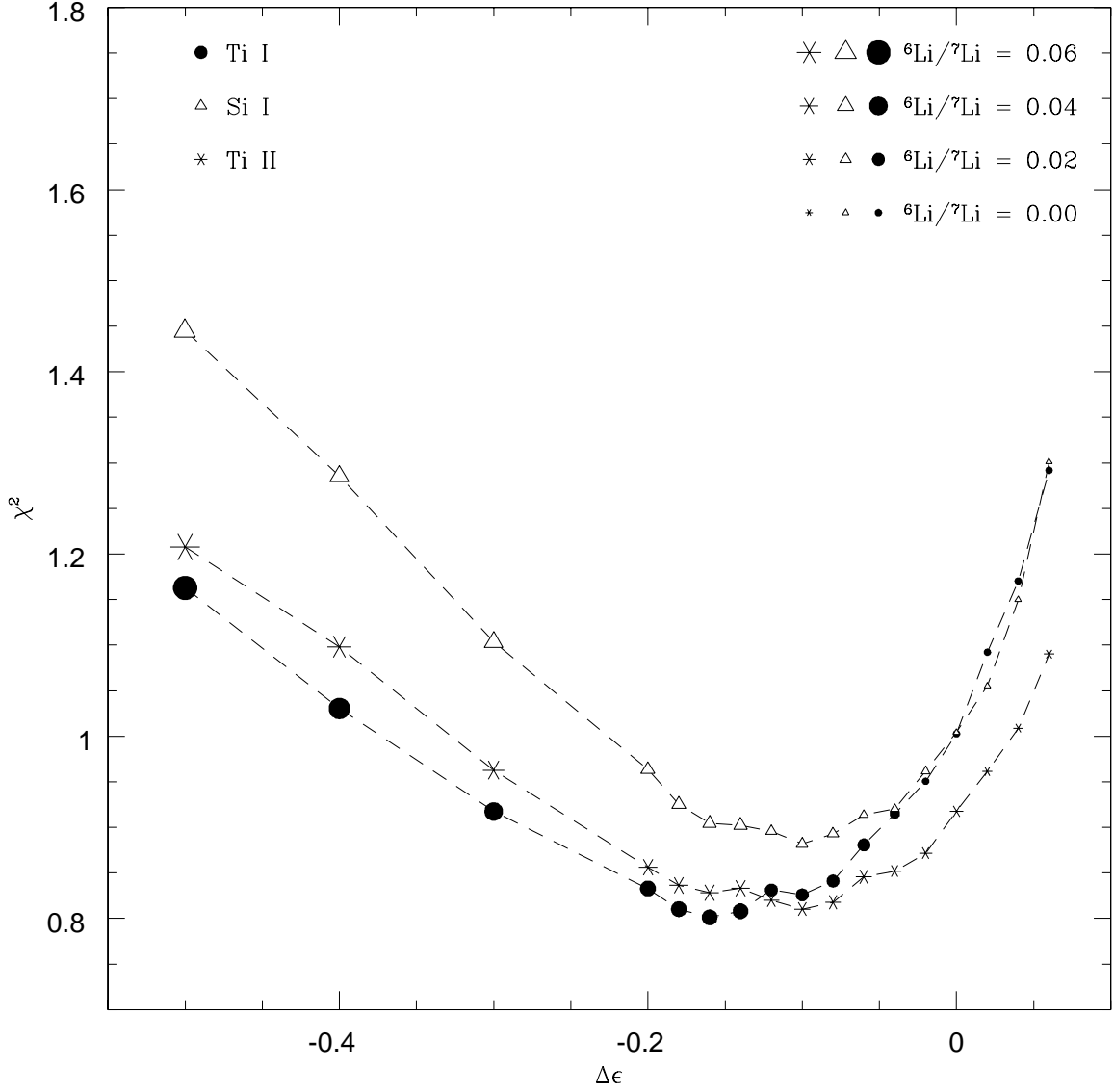}
\caption{Variations in $\chi^2_{min}$ for 47 Ursa
Majoris using different values of the abundance of the blending line
at 6708.025 \AA.  The $^6$Li/$^7$Li ratio is represented by the size
of the marker.  The best fit is found for a change in the abundance of
the blending element of $\sim-1.6$ for all of the blending options,
but the improvement in $\chi^2$ is minimal.\label{abtest}}
\end{figure}

\begin{figure}
\plotone{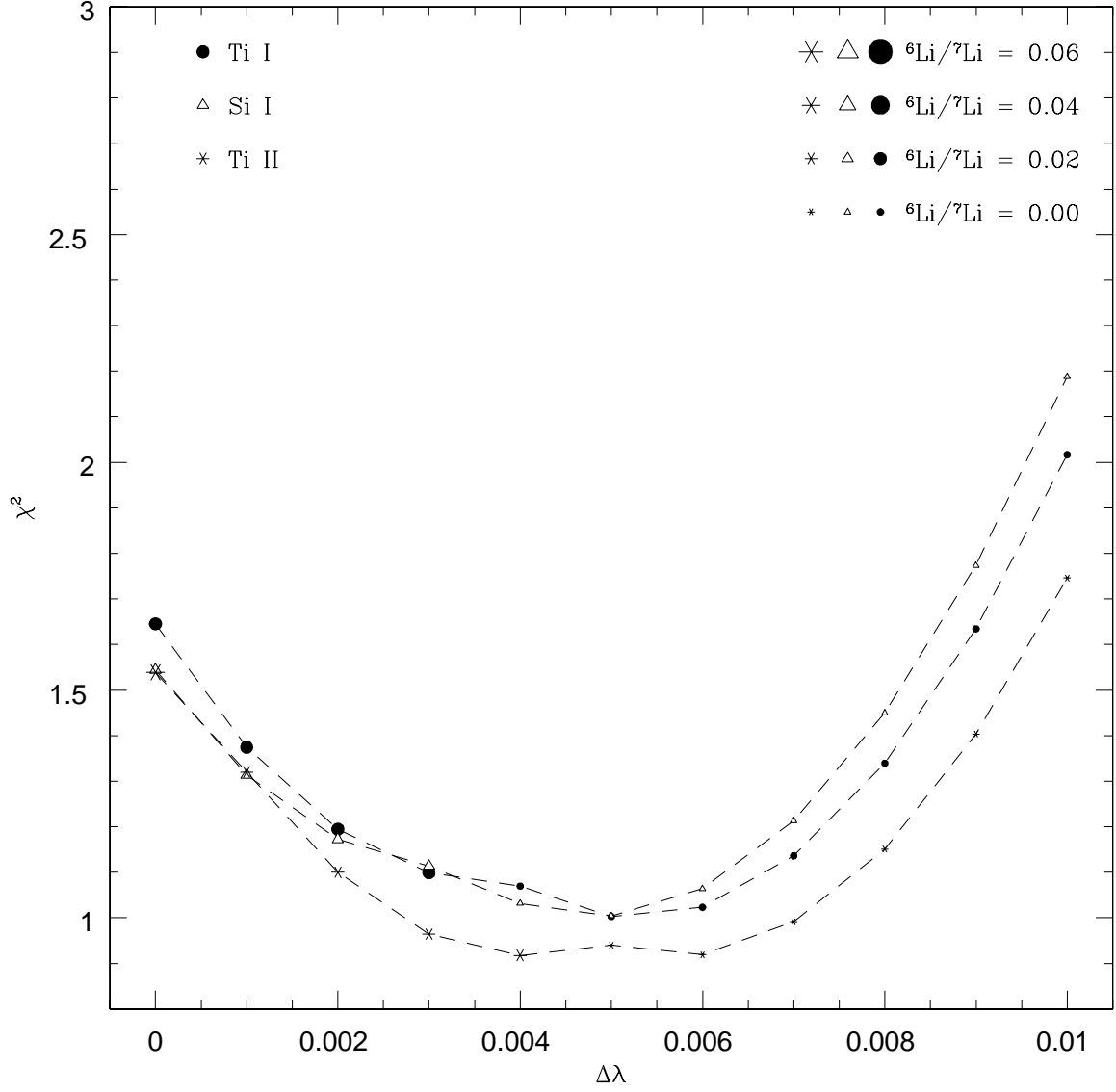}
\caption{Variations in $\chi^2_{min}$ for 47 Ursa
Majoris using different wavelength shifts for the lithium convective
shift.  The $^6$Li/$^7$Li ratio is represented by the size of the
marker.\label{wltest}}
\end{figure}

Adjustments in the wavelength shift and abundance of blending elements
were tested to measure the effects of uncertainties in both the
overall analysis as well as the line list.  Figure \ref{abtest} shows
the variation in $\chi^2_{min}$ with changes in the abundance of Ti
and Si from the values taken from the literature for 47 UMa, with the
lithium isotopic ratio represented by the size of the points.  As the
abundance of the blending element is decreased, the best fit value of
$^6$Li/$^7$Li increases in order to fit the line.  For most stars the
$\chi^2_{min}$ decreased slightly when the blending abundances were
decreased between 0.0 and 0.15 dex below the published
values. Decreasing the wavelength shift for the lithium line also
increased the $^6$Li/$^7$Li ratio needed to fit the line, though the
sensitivity of the isotopic ratio to the wavelength shift was less
than the sensitivity to the blending abundance (see Figure
\ref{wltest}). Considering the uncertainties in the line strength of
the blending line at 6708.025 \AA\ and the convective shifts, we
cannot rule out these variations in abundance and wavelength shift,
and therefore the $^6$Li/$^7$Li could be slightly higher than the
values found using abundances from the literature.  In Table
\ref{results}, the best-fit $^6$Li/$^7$Li value found with the
blending abundance and lithium wavelength shift kept as free
parameters is noted in parentheses next to the original value
calculated with the blending abundance held at literature values.

The results of the $\chi^2$ analyses for the three different choices
for the blending feature at 6708.025 \AA\ were very similar, with a
maximum difference in the best-fit value of 0.02 dex.  There did not
appear to be systematic differences between the quality of fits
achieved with the different line lists.  The best-fit values for the
lithium isotopic ratio for all the program stars were consistent with
a non-detection of $^6$Li within errors: none of the stars had
isotopic ratios above 0.0 for all three scenarios, and $\upsilon$ And
was the only star with an isotopic ratio above 0.0 for two of the
three line list scenarios.  The maximum best-fit isotopic ratio found
for any analysis was 0.02; the maximum 1- $\sigma$ upper limit was
0.05 for each of the two lithium-poor planet-bearing stars and 0.03
for HD 209458, the lithium-rich planet-bearing star.  The two
comparison stars had upper limits of 0.05 as well.

\section{Discussion} 
\label{discuss_sec}

The results found for all of the planet-bearing stars in this sample
are consistent with the results of R02, who found no $^6$Li present in
any of their program stars or their reference stars.  In particular,
the derived lithium ratios and upper limits for HD 209458 coincide
exactly.  R02 observed three stars with lithium abundances below 2.2
dex: HD 10697 (1.9 dex), HD 89744 (2.1 dex), and 16 Cyg B (0.7 dex).
Both HD 10697 and 16 Cyg B have effective temperatures around 5600K,
while HD 89744 has an effective temperature of 6338K.  Though no plots
of the $\chi^2$ analysis were published, upper limits for the isotopic
ratios of the three stars were 0.06, 0.03 and 0.03 respectively.
Though R02 only performed their analysis using the Ti I line at
6708.025 \AA, the fact the their results are concurrent with the
results from all of our analyzes suggests that these limits are
robust.  Currently the nature of the blending line at 6708.025 \AA\ is
still uncertain, and therefore no strong conclusions can be drawn from
the combined results of the lithium-poor stars in both studies, but it
appears that the evidence supports a lack of significant enhancement
of $^6$Li in most planet-bearing stars.

However, the value for the lithium isotopic ratio in HD 82943 is still
a mystery; the revised value of 0.05$\pm$0.02 reported by I03, who
used the Si I line in their analysis, is in conflict with the null
result reported by R02 using the Ti I lines.  But if the isotopic
ratio of HD 82943 is found to be definitively different from all other
stars studied to date, then a case may be made that this star has
undergone an unusual recent pollution event.

The upper limits on the lithium isotopic ratio can be translated into
an upper limit on the number of $^6$Li atoms in the convective zone of
a star.  Standard evolutionary models predict that the depth of the
convective mixing layer is relatively stable after a star reaches the
main sequence, and decreases with increasing stellar mass. The mass of
the convective envelope of 47 UMa, with a mass of 1.03$\Msun$ and an
effective temperature of 5800K, would be $\sim0.02\Msun$
\citep{pin01,mur01}, and a lithium abundance of log
$\epsilon(Li)=1.67$ and an isotopic ratio of 0.05 would be equal to
$\sim3.7\times10^{43}$ atoms of $^6$Li.  For a hotter, more massive
star such as $\upsilon$ And the convective envelope is between
0.001$\Msun$ and 0.01$\Msun$, depending on the age of the star and the
models used.  Using 0.005$\Msun$ as an estimate, a lithium abundance of
log $\epsilon(Li)=2.19$ and an isotopic ratio of 0.05 would also be
equal to $\sim3.7\times10^{43}$ atoms of $^6$Li.  Assuming an
abundance for protoplanetary material similar to the meteoric values
in our own solar system, one would expect $N(Li)/N(H)=2\times10^{-9}$
and $^6$Li/$^7$Li = 0.08, and a quantity of material with the mass of
Jupiter would have $\sim1.3\times10^{44}$ atoms of $^6$Li
\citep{and89}.  This suggests that $\sim0.3$ Jupiter masses of
material would have to fall onto a star fully depleted in $^6$Li to
immediately produce the upper limits on the isotopic ratio seen in
these two stars.  For HD 209458, the 1- $\sigma$ upper limit of 0.03
equates to $\sim8.9\times10^{43}$ $^6$Li atoms, which would be
equivalent to 0.7 Jupiter masses of material.

If circumstellar material was deposited at some time in the past, or
$^6$Li is not completely depleted in the PMS phase, the situation
becomes more complex.  Theoretical models of lithium depletion in the
pre-main sequence and main sequence phases \citep{pro89,pia02} of the
evolution of solar-type stars suggest that the amount of lithium
destroyed over the lifetime of a star depends critically on a variety
of parameters such as metallicity, rotational mixing, opacities and
the efficiency of convection.  However, it is generally agreed that
essentially all $^6$Li, and a large fraction of $^7$Li, is destroyed
in the PMS phase in stars with masses less than 1.4$\Msun$; for stars
above this mass the convection zone does not extend to temperatures
sufficient to destroy lithium efficiently.  Once the star reaches the
main sequence at an age of approximately 10 to 20 million years, the
convection zone retreats and classical models predict a cessation of
lithium destruction.  However, this prediction is at odds with
observations \citep{jon99} which show a decrease in average lithium
abundance with age on the main sequence.  Main sequence Li depletion
appears to be more severe in lower mass stars, except for a notable
increase in depletion for stars between 1.2 and 1.5$\Msun$
\citep{boe86}.  This phenomena is most likely due to rotational or
other macroscopic mixing processes that result in a larger surface
mixing layer \citep{mur01}, and most likely $^6$Li is destroyed at a
higher rate than $^7$Li similar to the timescales predicted for
pre-main sequence depletion.

Previous calculations of the ages of our program stars using a
combination of age indicators find ages of approximately 6.5 Gyr for
47 UMa and 3 Gyr for $\upsilon$ And \citep{lac99}, and R02 estimate an
age of 4 Gyr for HD 209458; this would suggest that any planetary
dynamics due to interactions with the protoplanetary disk that may
have deposited giant planets or massive amounts of planetary material
would most likely have occurred at much earlier times.  If any $^6$Li
was to be maintained in the convective envelopes of solar-mass stars
such as 47 UMa, it would most likely be greatly decreased from the
initial amount deposited onto the parent star; calculations by
\citet{mon02} suggest that $^6$Li would be depleted by more than an
order of magnitude over 5 Gyr.  However, higher-mass stars may be able
to maintain $^6$Li in their small convective zones for much longer,
though these timescales would be shortened for higher metallicity
stars.  This suggests that the upper limits on $\upsilon$ And may be
more meaningful than those calculated for 47 UMa, and the slight
possibility of a non-zero lithium ratio in $\upsilon$ And is not
completely unrealistic.

In order to reduce the uncertainties in the synthetic line synthesis,
several improvements in future observations and analysis can be
employed.  The most severe limitations on the accuracy of the isotopic
abundance analysis are the uncertainties in the identity and abundance
of the underlying blends around the $^6$Li lines and the uncertainties
in the line shape and wavelength shift due to unknown convective and
granular motions.  To improve the line identifications, additional
observations of metal-rich stars with a range of effective temperature
combined with laboratory measurements of metal lines in the lithium
region would allow a more complete assessment of the blending lines.
I03 have begun this process, but a larger sample of stars with varying
metallicities is needed.  To reduce the ambiguities in the line shape
and position, a more complex model of the stellar atmospheric
processes must be used.  Three-dimensional hydrodynamic models of
convection, though currently computationally expensive and restricted
in parameter space, are able to accurately predict the line shifts for
lines in the solar spectrum \citep{asp00}, and applications to non-LTE
calculations of lithium have recently been performed \citep{asp03}.
Two-component granulation models \citep{bor03} are also effective in
predicting line shifts and atomic parameters, and may be more
effective since they can be easily integrated in to line synthesis
applications.

\section{Conclusion} 
\label{concl_sec}

We have performed a search for $^6$Li in several planet-bearing stars
with low lithium abundances in order to investigate the possibility of
pollution due to infalling planets or planetary material and explore
the differences between line lists used in previous studies.  In order
to fit the nearby blending features in the lithium region, we have
enhanced previous atomic and molecular line lists using a combination
of lines from previously published line lists and semi- empirical fits
to the solar spectrum and a laboratory carbon arc spectrum.

Analyses of the lithium regions of the three planet-bearing stars 47
UMa, $\upsilon$ And, and HD 209458 using $\chi^2$ minimization find
lithium isotopic ratios consistent with the absence of a detectable
amount of $^6$Li.  1-$\sigma$ upper limits suggest that if any
planetary material was recently deposited on the host stars, it must
be less than half the mass of Jupiter.  However, if material was
deposited on the stars at much earlier times, destruction of the
fragile isotope could have rendered any pollution undetectable.

\acknowledgements We thank Chris Sneden, Guillermo Gonzalez, Chris
Churchill and Caylin Mendelowitz for their help with the reduction and
analysis, Robert Fields for his invaluable help in constructing the CN
line list, and Richard Wade and Steinn Sigurdsson for the general
knowledge they provided.  We also thank the referee for helpful
comments and suggestions.  This work was supported by NASA grants
NGA5-12115, NAG5-11427, and NAG5-13320, NSF grants AST-0138235 and
AST-0243090 and the Penn State Eberly College of Science.  The
Hobby-Eberly Telescope is operated by McDonald Observatory on behalf
of The University of Texas at Austin, the Pennsylvania State
University, Stanford University, Ludwig-Maximilians-Universit\"{a}t
M\"{u}nchen, and Georg-August-Universit\"{a}t G\"{o}ttingen.

\clearpage

\clearpage

\clearpage

\begin{figure}
\plotone{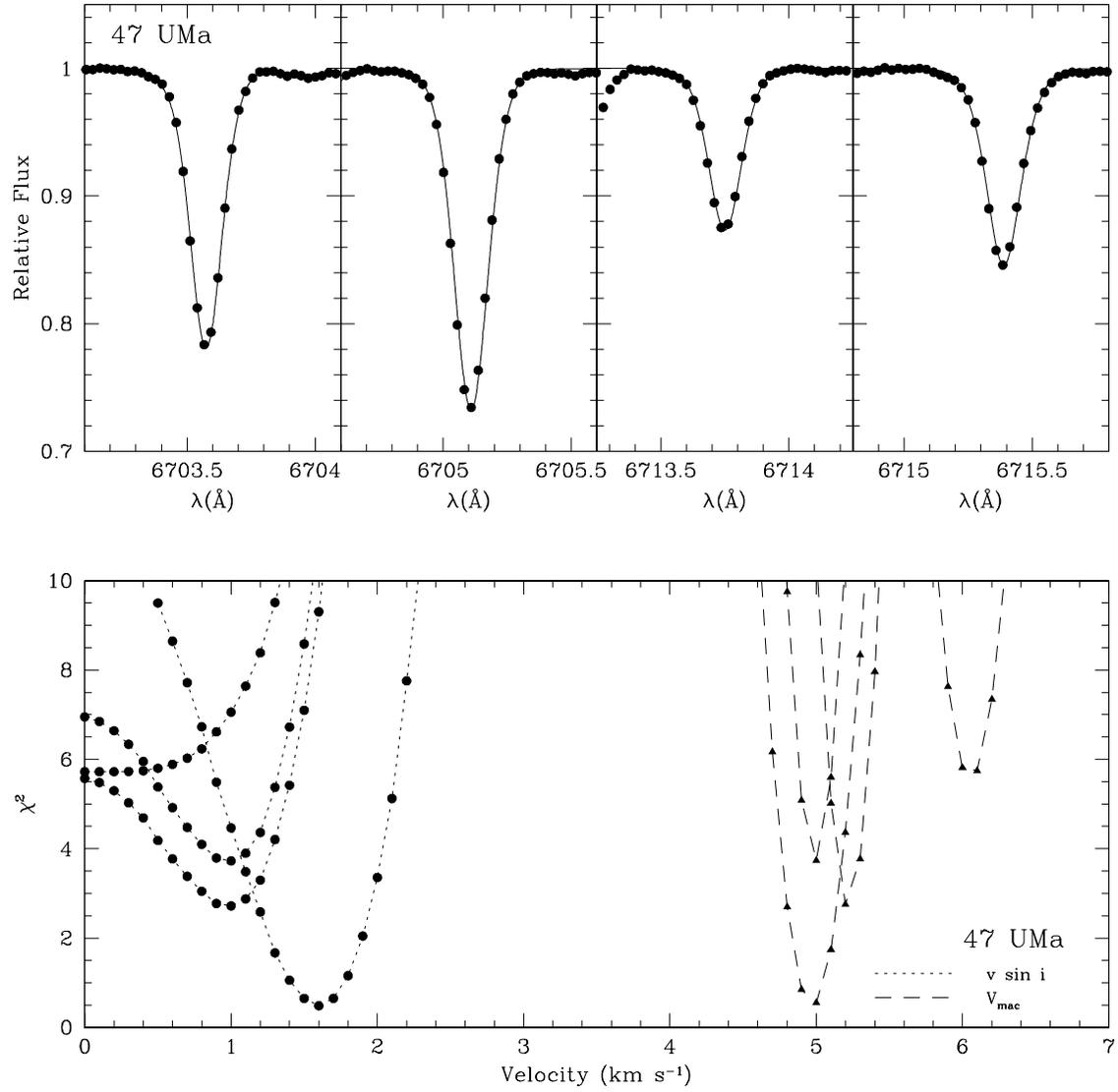}
\caption{Results of $\chi^2$ fits to nearby Fe I
lines for 47 UMa using adjustments in $v \sin i$ and
$V_{mac}$.  The best-fit values were found to within 0.01 km
$s^{-1}$.\label{broad}}
\end{figure}

\clearpage

\begin{figure}
\plotone{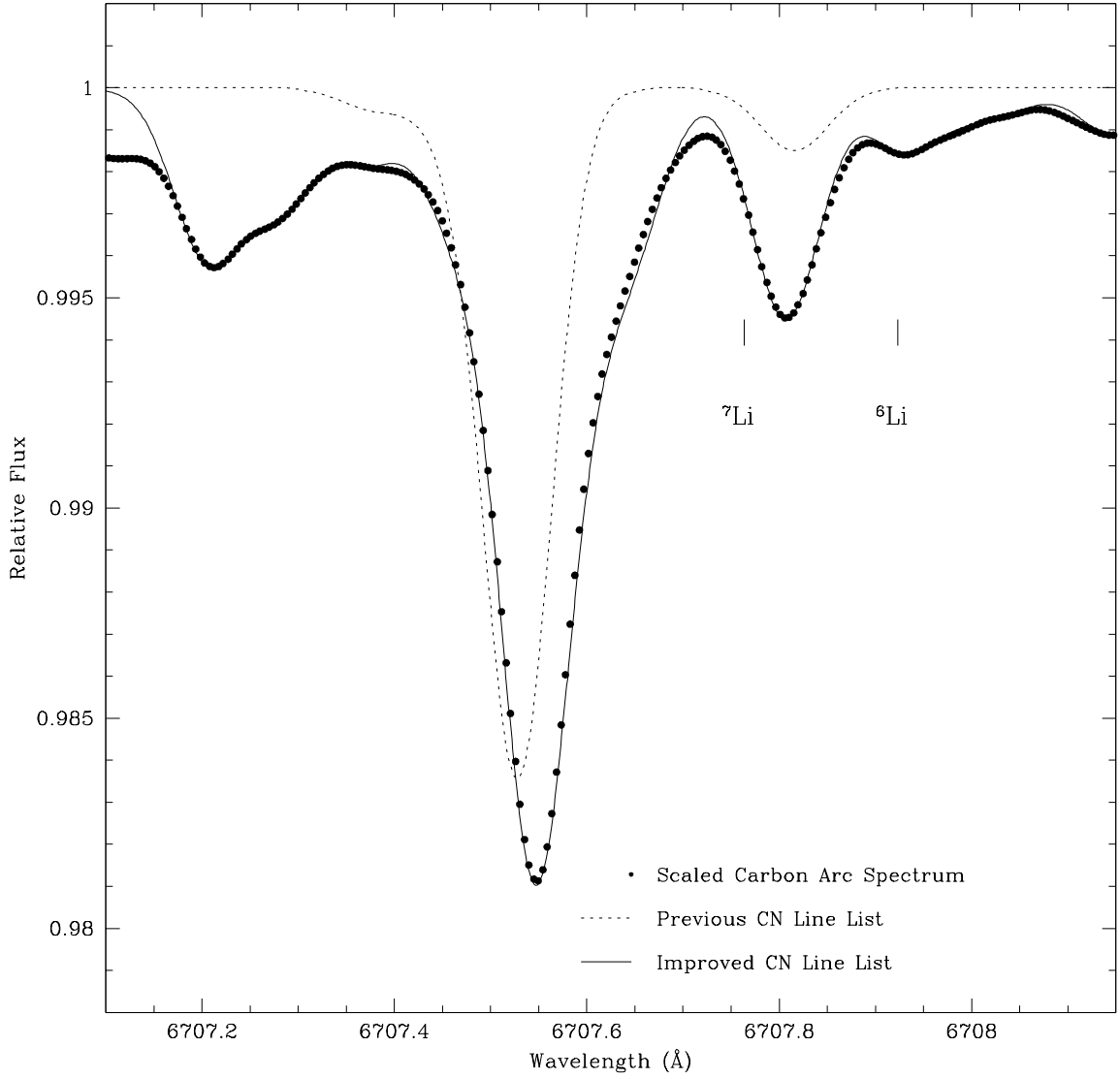}
\caption{Synthesized spectra for the CN line lists
used in this study and the list used in \citet{red02} overlayed on the
CN arc taken from the NSO database.  The improvement in fit can be
clearly seen.\label{cnarc}}
\end{figure}

\clearpage

\begin{figure}
\plotone{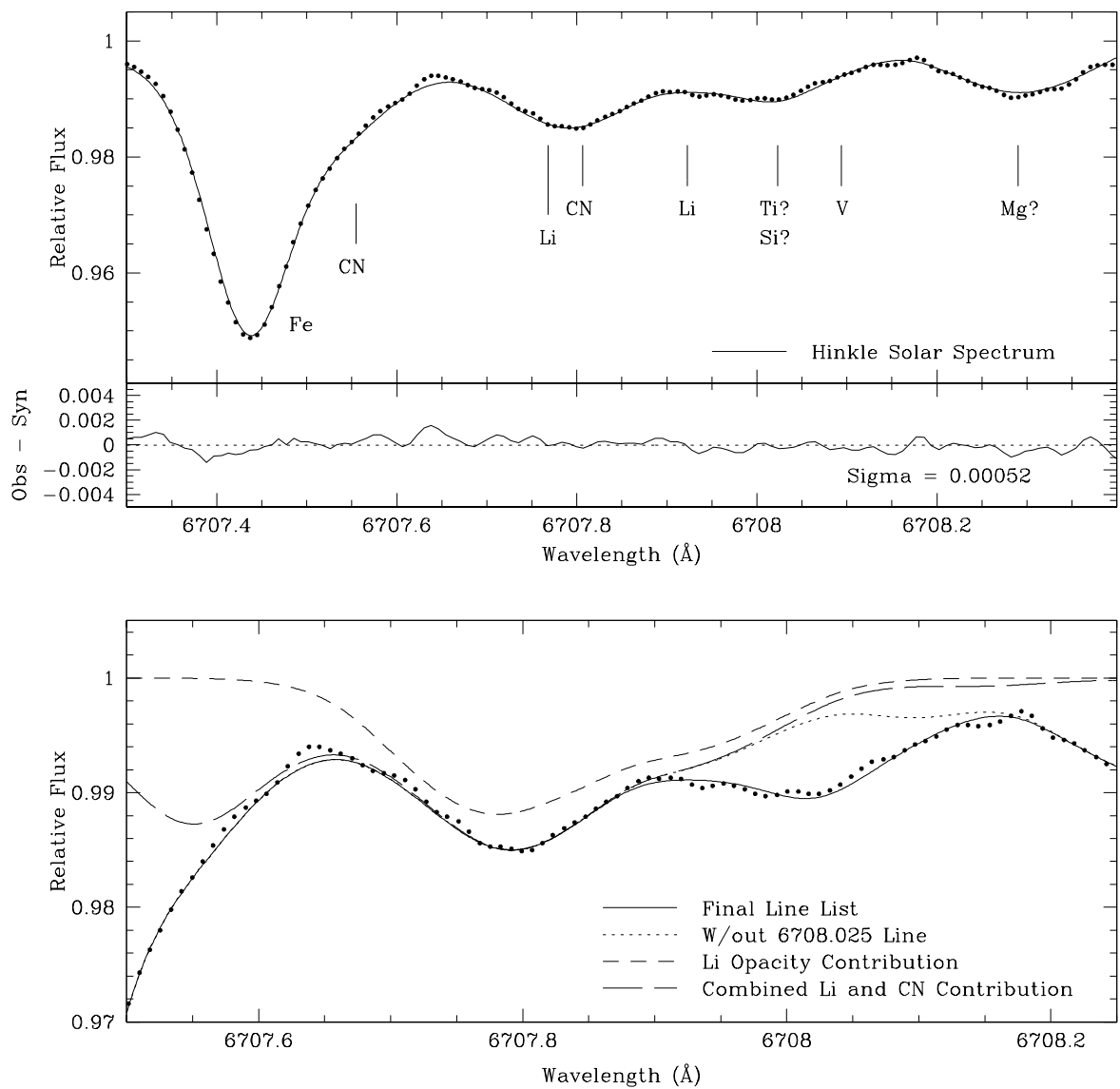}
\caption{Top panel: Synthesized spectra using the
final complete line list overlayed on the solar spectrum from
\citet{hin00}.  Bottom panel: Detailed view of the lithium doublet and
the blending lines.  Four synthetic spectra are plotted: the final
complete line list (solid), the full list minus the blending line at
6708.025 \AA\ (dotted), the contribution from lithium only (short
dashed), and the combined contribution from CN and lithium (long
dashed).  The importance of the red CN band in the region is apparent,
as are the blends near 6708 \AA.\label{solar}}
\end{figure}

\clearpage

\clearpage

\clearpage

\clearpage

\clearpage

\clearpage

\begin{deluxetable}{lcccccccccc}
\tabletypesize{\scriptsize} \tablewidth{0pt}
\tablecaption{Observational information and stellar characteristics
for the program stars taken from the literature.  Spectral types were
taken from the SIMBAD database.  The first reference is for the
atmospheric parameters, and the second reference is for the stellar
mass.\label{params}}
\tablehead{
\colhead{Star} & \colhead{V} & \colhead{S/N} & \colhead{$T_{eff}$} & 
\colhead{log g} & \colhead{$\xi_t$} & \colhead{[Fe/H]} & \colhead{Sp.Type} & 
\colhead{Mass} & \colhead{Age\tablenotemark{a}} &
\colhead{Ref\tablenotemark{b,c}}\\
\colhead{} & \colhead{} & \colhead{pixel$^{-1}$} & \colhead{(K)} &
\colhead{cm $s^{-2}$} & \colhead{km $s^{-1}$} & \colhead{} & \colhead{} & 
\colhead{$\Msun$} & \colhead{Gyr} & \colhead{}
}
\startdata
47 UMa	     & 5.10 & 1100 & 5800 & 4.25 & 1.00 &  0.01 & G0V  & 1.0 & 
 6.0-7.0 & 1,1\\
$\upsilon$ And & 4.09 &  900 & 6140 & 4.12 & 1.35 &  0.12 & F8V  & 1.3 &
 3.0-4.0 & 2,1\\ 
HD 209458    & 7.65 &  350 & 6063 & 4.38 & 1.02 &  0.04 & G0V  & 1.1 & 
 4.0-5.0 & 3,1\\ 
HD 22484     & 4.28 &  300 & 5950 & 4.00 & 1.35 & -0.11 & F9IV & 1.1 & 
 6.0-7.0 & 4,2\\ 
HD 136064    & 5.10 &  850 & 6077 & 3.94 & 1.52 & -0.02 & F9IV & 1.3 & 
 3.0-5.0 & 5,2\\
\enddata
\tablenotetext{a}{Stellar age ranges compiled from
\citet{ibu02}, \citet{lac99}, and \citet{ng98}}
\tablenotetext{b}{References for stellar atmospheric parameters- 
1: \citet{gon98}; 2: \citet{gon00}; 3:\citet{gon01}; 4: \citet{gra96}; 
5: \citet{car00}}
\tablenotetext{c}{References for stellar masses- 1: \citet{sch03}; 2:
\citet{ng98}}
\end{deluxetable}

\clearpage

\begin{deluxetable}{lcccccccc}
\tabletypesize{\scriptsize}
\tablewidth{0pt}
\tablecaption{Results from $\chi^2$ tests for the program stars.
Final values for $v \sin i$ and $V_{mac}$ are given, and the three
different columns for $^6$Li/$^7$Li are for the three different
choices for the blending line at 6708.025 \AA.  Values in parentheses
were calculated with blending abundance and lithium convective shifts
kept as free parameters.\label{results}}
\tablehead{
\colhead{Star} & \colhead{$v \sin i$} & \colhead{$V_{m}$} & 
\colhead{log $\epsilon$(Li)} & \colhead{$^6$Li/$^7$Li (Ti I)} & 
\colhead{$^6$Li/$^7$Li (Si I)} & \colhead{$^6$Li/$^7$Li (Ti II)}
}
\startdata
47 UMa         & 1.19 & 5.06 & 1.67$\pm$0.02 
   & 0.00$\pm$0.04 (0.03) & 0.00$\pm$0.04 (0.02) & 0.02$\pm$0.03 (0.03)\\
$\upsilon$ And   & 8.33 & 7.88 & 2.19$\pm$0.02 
   & 0.02$\pm$0.03 (0.02) & 0.00$\pm$0.03 (0.00) & 0.01$\pm$0.03 (0.02)\\ 
HD 209458      & 4.20 & 5.23 & 2.64$\pm$0.01 
   & 0.00$\pm$0.03 (0.00) & 0.00$\pm$0.03 (0.00) & 0.00$\pm$0.03 (0.00)\\
HD 22484       & 2.17 & 5.84 & 2.43$\pm$0.03 
   & 0.00$\pm$0.05 (0.00) & 0.00$\pm$0.05 (0.00) & 0.00$\pm$0.05 (0.00)\\
HD 136064      & 3.33 & 6.88 & 1.81$\pm$0.02
   & 0.02$\pm$0.05 (0.02) & 0.00$\pm$0.05 (0.00) & 0.00$\pm$0.05 (0.00)\\
\enddata
\end{deluxetable}

\clearpage

\begin{deluxetable}{lllll}
\tabletypesize{\scriptsize}
\tablewidth{0pt}
\tablecaption{Final CN line list used for this study.  Original line
identifications and parameters were collected from the references
below, and modified to fit the NSO carbon arc spectrum.\label{cn}}
\tablehead{
\colhead{Identification} & \colhead{Wavelength(air)} & \colhead{LEP} &
\colhead{log $gf$} & \colhead{Ref\tablenotemark{a}}\\
\colhead{} & \colhead{(\AA)} & \colhead{($eV$)} & \colhead{($dex$)} &
\colhead{}}
\startdata
$Q_{1}(43)(6,2)   $ & 6702.5314 & 0.94 & -1.608 & J\\ 
$Q_{12}(16)(7,3)  $ & 6703.9421 & 0.81 & -1.869 & J\\
$P_{1}(16)(7,3)   $ & 6704.0157 & 0.81 & -2.009 & J\\ 
$Q_{1}(96)(6,1)   $ & 6705.8851 & 2.35 & -1.792 & J\\
$P_{12}(19)(12,7) $ & 6705.8883 & 1.79 & -3.298 & J\\ 
$R_{21}(39)(7,3)  $ & 6705.8905 & 1.10 & -3.227 & J\\
$P_{1}(23)(12,7)  $ & 6705.9816 & 1.82 & -1.890 & K\\ 
$Q_{2}(93)(11,5)  $ & 6706.5476 & 3.13 & -1.359 & J\\
$Q_{2}(80)(8,3)   $ & 6706.5665 & 2.19 & -1.650 & J\\ 
$R_{12}(22)(7,3)  $ & 6706.6568 & 0.87 & -3.001 & K\\
$Q_{1}(22)(7,3)   $ & 6706.7329 & 0.87 & -1.807 & K\\ 
$R_{1}(34)(12,7)  $ & 6706.8440 & 1.96 & -2.775 & K\\
$P_{2}(83)(7,2)   $ & 6706.8626 & 2.07 & -1.882 & J\\ 
$Q_{2}(47)(11,6)  $ & 6707.2052 & 1.97 & -1.222 & K\\
$Q_{2}(60)(10,5)  $ & 6707.2823 & 2.04 & -1.333 & J\\ 
$Q_{1}(85)(12,6)  $ & 6707.3706 & 3.05 & -0.522 & J\\
$P_{12}(13)(7,3)  $ & 6707.4569 & 0.79 & -3.055 & K\\ 
$Q_{1}(28)(12,7)  $ & 6707.4695 & 1.88 & -1.581 & K\\
$Q_{2}(44)(6,2)   $ & 6707.5450 & 0.96 & -1.598 & K\\ 
$Q_{2}(29)(12,7)  $ & 6707.5947 & 1.89 & -1.451 & K\\
$P_{21}(44)(6,2)  $ & 6707.6453 & 0.96 & -2.310 & K\\ 
$R_{1}(64)(5,1)   $ & 6707.8071 & 1.21 & -1.853 & K\\
$R_{1}(61)(19,12) $ & 6707.8475 & 3.60 & -2.417 & J\\ 
$P_{2}(39)(20,13) $ & 6707.8992 & 3.36 & -3.110 & J\\
$Q_{21}(35)(12,7) $ & 6707.9300 & 1.98 & -1.651 & J\\ 
$R_{21}(72)(10,5) $ & 6707.9800 & 2.39 & -2.027 & K\\
$R_{2}(35)(12,7)  $ & 6708.0261 & 1.98 & -2.031 & J\\ 
$P_{2}(42)(11,6)  $ & 6708.1470 & 1.87 & -1.884 & J\\
$Q_{12}(12)(17,11)$ & 6708.3146 & 2.64 & -1.719 & J\\ 
$P_{1}(12)(17,11) $ & 6708.3700 & 2.64 & -2.540 & J\\
$P_{12}(33)(6,2)  $ & 6708.4200 & 0.76 & -3.440 & J\\ 
$R_{1}(94)(7,2)   $ & 6708.5407 & 2.50 & -1.876 & J\\
$P_{2}(42)(11,6)  $ & 6708.6345 & 1.87 & -1.584 & K\\
\enddata
\tablenotetext{a}{Reference for original wavelength- J:
\citet{jor90}; K: \citet{kot80}}
\end{deluxetable}

\clearpage

\begin{deluxetable}{lllll}
\tabletypesize{\scriptsize}
\tablewidth{0pt}
\tablecaption{Final atomic line list in the vicinity of the lithium
lines.  Lines marked with an asterisk have been added for this
study.\label{atoms}}
\tablehead{
\colhead{Wavelength} & \colhead{Element} & \colhead{LEP} 
& \colhead{log $gf$} & \colhead{Ref\tablenotemark{a}}\\
\colhead{(\AA)} & \colhead{} & \colhead{($eV$)} & \colhead{($dex$)} & 
\colhead{}
}
\startdata
6706.8800  & Fe II  & 5.96 & -4.100 & V \\
6707.0100  & Si I   & 5.95 & -2.640 & V \\
6707.4330  & Fe I   & 4.61 & -2.320 & R \\
6707.4730  & Sm II  & 0.93 & -1.777 & V \\ 
6707.5180  & V I    & 2.74 & -1.995 & V \\ 
6707.6440  & Cr I   & 4.21 & -2.667 & V \\ 
6707.7400  & Ce II  & 0.50 & -3.810 & R \\
6707.7520  & Sc I   & 4.05 & -2.654 & V \\
6707.7561  & $^7$Li & 0.00 & -0.428 & H \\ 
6707.7682  & $^7$Li & 0.00 & -0.206 & H \\ 
6707.7710  & Ca I   & 5.80 & -4.015 & R \\
6707.9066  & $^7$Li & 0.00 & -1.509 & H \\ 
6707.9080  & $^7$Li & 0.00 & -0.807 & H \\ 
6707.9187  & $^7$Li & 0.00 & -0.807 & H \\ 
6707.9196  & $^6$Li & 0.00 & -0.479 & H \\ 
6707.9200  & $^7$Li & 0.00 & -0.807 & H \\ 
6707.9230  & $^6$Li & 0.00 & -0.178 & H \\ 
6707.9640  & Ti I   & 1.88 & -6.903 & V \\
6708.0220* & Ti II  & 1.88 & -4.230 & M*\\
6708.0230* & Si I   & 6.00 & -2.890 & I*\\ 
6708.0250* & Ti I   & 1.88 & -2.202 & R*\\ 
6708.0728  & $^6$Li & 0.00 & -0.303 & H \\ 
6708.0940  & V I    & 1.22 & -2.313 & V \\
6708.2750  & Ca I   & 5.88 & -6.097 & V \\
6708.2900* & Mg I   & 5.75 & -2.678 & M*\\
6708.6090  & Fe I   & 5.45 & -3.405 & V \\
\enddata
\tablenotetext{a}{Reference for original data- V: VALD-2 database,; K: Kurucz 
database; R: \citet{red02}; I: \citet{isr03}; M: this study; H: \citet{hob99}}
\end{deluxetable}

\clearpage

\begin{deluxetable}{llll}
\tabletypesize{\scriptsize}
\tablewidth{0pt}
\tablecaption{Ti and Si lines used to check abundances for the
program stars taken from the literature.\label{checklines}}
\tablehead{
\colhead{Wavelength} & \colhead{Element} & \colhead{LEP} 
& \colhead{log $gf$}\\
\colhead{(\AA)} & \colhead{} & \colhead{($eV$)} & \colhead{($dex$)}}
\startdata
6680.1370 & Ti II & 3.10 & -2.075\\
6689.2480 & Ti I  & 2.25 & -1.602\\
6716.6730 & Ti I  & 2.49 & -1.143\\
 & & & \\
6721.8430 & Si I & 5.86 & -1.190\\
6739.5230 & Si I & 5.96 & -1.660\\
6741.6280 & Si I & 5.98 & -1.680\\
\enddata
\end{deluxetable}


\begin{thebibliography}

\bibitem[Asplund {\sl et al.}(2003)]{asp03} {Asplund, M., Carlsson,
	M., \& Botnen, A. V. 2003, A\&A, 399, L31}
\bibitem[Asplund {\sl et al.}(2000)]{asp00} {Asplund, M., Nordlund,
	\AA, Trampedach, R., Allende Prieto, C., and Stein, R.F. 2002, A\&A,
	359, 729}
\bibitem[Allende Prieto \& Garc\'{i}a L\'{o}pez(1998)]{all98} {Allende
	Prieto, C. \& Garc\'{i}a L\'{o}pez, R. J. 1998, A\&AS, 129, 41}
\bibitem[Allende Prieto {\sl et al.}(2002)]{all02} {Allende Prieto, C.,
	Lambert, D. L., Tull, R.G., \& MacQueen, P. J. 2002, ApJL, 566,L93}
\bibitem[Anders \& Grevesse(1989)]{and89} {Anders, E. \& Grevesse,
	N. 1989, Geochim. et Cosmochim Acta, 53, 197}
\bibitem[Boesgaard \& Tripicco(1986)]{boe86} {Boesgaard, A. M. \&
	Tripicco, M. J.  1986, ApJ, 302, L49}
\bibitem[Borrero {\sl et al.}(2003)]{bor03} {Borrero, J.M., Bellot
	Rubio, L.R., Barklem, P.S., \& del Toro Iniesta, J.C. 2003, A\&A, 404,
	749}
\bibitem[Brault \& M\"{u}ller(1975)]{bra75} {Brault, J. W. \&
	M\"{u}ller, E. A. 1975, SoPh, 41, 43}
\bibitem[Carretta {\sl et al.}(2000)]{car00} {Carretta, E., Gratton, R. G.,
	\& Sneden, C. 2000, A\&A, 356, 238}
\bibitem[Davis(1987)]{dav87} {Davis, S. P. 1987, PASP, 99, 1105}
\bibitem[de Laverny \& Gustafsson(1998)]{del98} {de Laverny, P. \&
	Gustafsson, B. 1998, A\&A, 332, 661}
\bibitem[Denn {\sl et al.}(1991)]{den91} {Denn, G. R., Luck, R. E., \&
	Lambert, D. L. 1991, ApJ, 377, 657}
\bibitem[Forestini(1994)]{for94} {Forestini, M. 1994, A\&A, 285, 473}
\bibitem[Gonzalez(1998)]{gon98} {Gonzalez, G. 1998, A\&A, 334, 221}
\bibitem[Gonzalez \& Laws(2000)]{gon00} {Gonzalez, G. \& Laws,
	C. 2000, AJ, 119, 390}
\bibitem[Gonzalez {\sl et al.}(2001)]{gon01} {Gonzalez, G., Laws, C., Tyagi,
	S., \& Reddy, B. E. 2001, AJ, 121, 432}
\bibitem[Gratton {\sl et al.}(1996)]{gra96} {Gratton, R. G., Carreta, E. \&
	Castelli, F. 1996, A\&A, 314, 191}
\bibitem[Grevesse(1968)]{gre68} {Grevesse, N. 1968, SoPh, 5, 159}
\bibitem[Hill(2003)]{hil03} {Hill, F. 2003, The National Solar
	Observatory Digital Library, http://diglib.nso.edu}
\bibitem[Hinkle {\sl et al.}(2000)]{hin00} {Hinkle, K., Wallace, L.,
	Valenti, J., \& Harmer, D. 2000 Visible and Near Infrared
	Atlas of the Arcturus Spectrum 3727 - 9300 \AA\ (San Francisco: ASP)}
\bibitem[Hobbs {\sl et al.}(1999)]{hob99} {Hobbs, L. M., Thorburn, J. A., \&
	Rebull, L. M. 1999, ApJ, 523, 797}
\bibitem[Ibukiyama \& Arimoto(2002)]{ibu02} {Ibukiyama, A. \& Arimoto,
	N. 2002, A\&A, 394, 927}
\bibitem[Israelian {\sl et al.}(2003)]{isr03} {Israelian, F., Santos, N.,
	Mayor, M., \& Rebolo, R. 2003, astro-ph/0304358}
\bibitem[Israelian {\sl et al.}(2001)]{isr01} {Israelian, G., Santos, N.C.,
	Mayor, M., \& Rebolo, R. 2001, Nature, 411, 163}
\bibitem[Jones {\sl et al.}(1999)]{jon99} {Jones, B. F., Fischer,
	D. \& Soderblom, D. R. 1999, AJ, 117, 117, 330}
\bibitem[Jorgensen \& Larsson(1990)]{jor90} {Jorgensen, U. G. \&
	Larsson, M. 1990, A\&A, 238, 424}
\bibitem[King {\sl et al.}(1997)]{kin97} {King, J. R., Deliyannis, C. P.,
	Hiltgen, D. D., Stephens, A., Cunha, K., \& Boesgaard, A. M. 1997, AJ,
	113, 1871}
\bibitem[Kotler {\sl et al.}(1980)]{kot80} {Kotlar, A. J., Field, R. W., \&
	Steinfeld, J. I. 1980, J.Mol.Spectr., 80, 86}
\bibitem[Kupka {\sl et al.}(1999)]{kup99} {Kupka, F., Piskunov, N.,
	Ryabchikova, T., Stempels, H., \& Weiss, W.W. 1999, A\&AS, 138, 119}
\bibitem[Kurucz(1995)]{kur95} {Kurucz, R. 1995: http://cfaku5.harvard.edu}
\bibitem[Lambert {\sl et al.}(1993)]{lam93} {Lambert, D. L., Smith, V. V.,
	\& Heath, J. 1993, PASP, 105, 568}
\bibitem[Lachaume {\sl et al.}(1999)]{lac99} {Lachaume, R., Dominik, C.,
	Lanz, T., \& Habing, H. J. 1999, A\&A, 348, 897}
\bibitem[Lin {\sl et al.}(1996)]{lin96} {Lin, D. N. C., Bodenheimer, P.,
	\& Richardson, D. C. 1996, Nature, 380, 606}
\bibitem[Montalb\'{a}n \& Rebolo(2002)]{mon02} {Montalb\'{a}n, J. \&
	Rebolo, R. 2002, A\&A, 386, 1039}
\bibitem[M\"{u}ller {\sl et al.}(1975)]{mul75} {M\"{u}ller, E.A., Peytremann,
	E. \& De La Reza, R. 1975, Solar Physics, 41, 53}
\bibitem[Murray {\sl et al.}(2001)]{mur01} {Murray, N, Chaboyer, B.,
	Arras, P., Hansen, B., \& Noyes, R.W. 2001, ApJ, 555, 801}
\bibitem[Murray {\sl et al.}(1998)]{mur98} {Murray, N., Hansen, B., Holman, M.,
	\& Tremaine, S. 1998, Science, 279, 69}
\bibitem[Nave {\sl et al.}(1994)]{nav94}{Nave, G., Johansson, S., Learner,
	R.C.M., Thorne, A.P., \& Brault, J.W. 1994, ApJS, 94, 221}
\bibitem[Ng \& Bertelli(1998)]{ng98} {Ng, Y. K. \& Bertalli, G. 1998,
	A\&A, 329, 943}
\bibitem[Piau \& Turck-Chi\`{e}ze(2002)]{pia02} {Piau, L. \&
	Turck-Chi\`{e}ze, S. 2002, ApJ, 566, 419}
\bibitem[Pinsonneault {\sl et al.}(2001)]{pin01} {Pinsonneault, M. H.,
	DePoy, D. L., \& Coffee, M. 2001, ApJ, 556, L59}
\bibitem[Proffitt \& Michaud(1989)]{pro89} {Proffitt, C. R., \&
	Michaud, G. 1989, ApJ, 346, 976}
\bibitem[Rasio \& Ford(1996)]{ras96} {Rasio, F. A. \& Ford,
	E. B. 1996, Science, 274, 954}
\bibitem[Reddy {\sl et al.}(2002)]{red02} {Reddy, B., Lambert, D., Laws, C.,
	Gonzalez, G. \& Covery, K. 2002, MNRAS, 335, 1005}
\bibitem[Sandquist {\sl et al.}(2002)]{san02} {Sandquist, E. L., Dokter,
	J. J., Lin, D. N. C.; Mardling, R. A. 2002, 572, 1012}
\bibitem[Schneider(2003)]{sch03} Schneider, J.  2003, The Extrasolar
	Planets Encyclopedia (Paris: Obs. Paris),
	http://cfa-www.harvard.edu/planets/encycl.html
\bibitem[Sneden(1973)]{sne73} {Sneden, C. 1973, Ph.D. thesis,
	Univ. Texas-Austin}
\bibitem[Tull(1998)]{tul98} {Tull, R. G. 1998, Proc. SPIE, 3355, 387}
\bibitem[Weidenschilling \& Marzari(1996)]{wei96} {Weidenschilling,
	S. J. \& Marzari, F. 1996, Nature, 384, 619}

\end{thebibliography}
\end{document}